\documentclass[conference]{IEEEtran}
\IEEEoverridecommandlockouts
\usepackage[colorlinks=true,linkcolor=black,citecolor=black]{hyperref}%

\usepackage{cite}
 \usepackage{colortbl}
\usepackage{amsmath,amssymb,amsfonts}
\usepackage{algorithmic}
\usepackage{graphicx}
\usepackage{textcomp}
\usepackage{enumitem}
\usepackage{xcolor}

\def\BibTeX{{\rm B\kern-.05em{\sc i\kern-.025em b}\kern-.08em
    T\kern-.1667em\lower.7ex\hbox{E}\kern-.125emX}}
\begin{document}

\title{A Review of Federated Learning in Energy Systems}

\author{
        \IEEEauthorblockN{Xu Cheng}
        \IEEEauthorblockA{\textit{Section of Energy Markets} \\
            \textit{Smart Innovation Norway}\\
            Halden, Norway \\
            xu.cheng@ieee.org}
    \and
        \IEEEauthorblockN{Chendan Li}
        \IEEEauthorblockA{\textit{Department of Marine Technology} \\
            \textit{Norwegian University of Science and Technology}\\
            Trondheim, Norway \\
            chendan.li@ntnu.no}
    \and
        \IEEEauthorblockN{Xiufeng Liu}
        \IEEEauthorblockA{\textit{Department of Technology,} \\
            \textit{Management and Economics} \\
            \textit{Technical University of Denmark}\\
            Produktionstorvet, Denmark \\
            xiuli@dtu.dk}
}

\maketitle

\begin{abstract}
With increasing concerns for data privacy and ownership,  recent years have witnessed a paradigm shift in machine learning (ML). An emerging paradigm, federated learning (FL), has gained great attention and has become a novel design for machine learning implementations. FL enables the ML model training at data silos under the coordination of a central server, eliminating communication overhead and without sharing raw data. In this paper, we conduct a review of the FL paradigm and, in particular, compare the types, the network structures, and the global model aggregation methods. Then, we conducted a comprehensive review of FL applications in the energy domain (refer to the smart grid in this paper). We provide a thematic classification of FL to address a variety of energy-related problems, including demand response, identification, prediction, and federated optimizations. We describe the taxonomy in detail and conclude with a discussion of various aspects, including challenges, opportunities, and limitations in its energy informatics applications, such as energy system modeling and design, privacy, and evolution.
\end{abstract}

\begin{IEEEkeywords}
Review, Federated Learning, Energy sector, Distributed learning, Privacy and security
\end{IEEEkeywords}

\section{Introduction}
Today the demand for energy is increasing rapidly in the global, due to population and economic growth. The building sector accounts for 40\% of total energy consumption and 60\% of electricity consumption \cite{li2019review}. Distributed energy resources are a viable solution to address this problem by integrating distributed energy systems at users. Distributed energy systems have the characteristics of high efficiency, low loss, low pollution, and flexible operations. On the other hand, for electricity supply, smart grids have intensively adopted the Internet of Things (IoT), smart meters, and advanced communication networks (e.g. 5G) and data management systems, which form an advanced energy management infrastructure. Smart grids and distributed energy systems together form a giant energy of the Internet (IoE) to provide stable and sustainable energy to end users. However, several prominent issues arise with the provision of various energy services, the management of massive data, and the offloading of computing workloads from the cloud. First, data privacy becomes a major concern as customer data is often used to train data-driven models, e.g., for detecting malicious nodes in IoE.  Therefore, it has to access private customer information, such as consumption profiles and habits, to train the model. Second, data owners may not be willing to share their data to a centralized server for model training, mainly due to privacy, commercial competition, and technical barrier reasons \cite{cheng2022class}. At the legal level, governments or organizations around the world are also increasingly committed to data privacy protection. For example, the European Union has enforced the General Data Protection Regulation (GDPR) since 2017 \cite{voigt2017eu}. Third, the computing resource requirements for traditional centralized model training are relatively high due to the management of large amounts of data sets. Last, data transmission over the network is vulnerable to cyberattacks and introduces overhead. Therefore, there is a pressing need for a feasible solution to address these issues. 

Federated learning (FL) is an ideal paradigm, which was first introduced by Google in 2016 \cite{konevcny2016federated}. FL is essentially a distributed learning approach, where multiple distributed clients train machine learning models separately using their own local data under the coordination of a central server, then the server aggregates the trained models into a final global model. Training is an iterative process that involves updating local models and aggregating the global model until the model converges or the training reaches the predefined number of interactions. FL is a major shift from centralized and expensive machine learning to a distributed manner that can use many distributed computing resources. This learning paradigm improves data privacy, as the raw data remain on the local device, eliminating network overhead as only the model updates are exchanged. In recent years, FL has been used successfully in industries related to IoT and edge computing, mainly for privacy protection, but it is attracting more and more interest.  Some literature reviews were performed not only on FL technology itself, such as the different FL architectures and their learning settings, e.g., \cite{litian2020federated,yin2021comprehensive,zhang2021survey,mothukuri2021survey,nguyen2021federated,rahman2021challenges}, but also its applications, e.g., \cite{li2020review,khokhar2022review,antunes2022federated,ali2022federated,liu2021federated}. However, there is still no systematic review of FL applications in energy. This is likely due to more research efforts in areas directly related to the protection of personal information privacy, such as medicine, healthcare, and insurance. On the contrary, the literature on FL in energy is much fewer and all of them emerged within the last two years. However, we have seen that growth is fast and its applications go beyond addressing data privacy, which is raising great research attention in the energy community. Therefore, we believe that it is necessary to conduct a systematic review. In this paper,  we will classify and summarize the different solutions based on FL and their applications in the energy sector. Moreover, unlike other reviews of FL technologies which focus on different architectures and complementary technologies to enhance FL, we pay special attention to different aggregate algorithms to generate the global model. This review can help advance understanding of FL technology and promote its use cases in the energy sector.

In summary, this paper makes the following contributions.
\begin{itemize}
    %\item Taxonomy of the FL technology has been listed in the introduction of the FL concept, the main genres of the FL technology framework will be introduction based on it data partition method and network architecture. Special discussion on the aggregation method is give for main stream FL frameworks.
     \item We list the taxonomy of FL technology, classify FL frameworks according to data partitioning methods and network structures, and, in particular, summarize the mainstream aggregation algorithms for generating the global model.
    \item We review FL use cases in the smart grid domain, identify their technological features, and provide an overview of current FL in the energy domain.
    \item We discuss the challenges and opportunities for the FL technology itself and for applications in energy and suggest possible solutions to shed light on the wider adoption of FL in energy.
    %The current challenges of the FL technology applied on the energy are identified and possible solution for the future application are suggested with the hope to shed light on the future adoption of FL in the wider energy sector. 
\end{itemize}

The remainder of this paper is structured as follows. Section~\ref{sec:fl} gives an overview of the FL paradigm, and surveys the methods for global model generation; Section~\ref{sec:usecases} reviews the use cases of FL in energy systems; Section~\ref{sec:challenges} discusses the challenges, opportunities, and limitations of FL in energy; Section~\ref{sec:con} concludes the paper and presents the future research directions.

\section{Federated Learning Concept}
\label{sec:fl}

\subsection{Overview}

FL provides a novel machine learning solution with distributed and enhanced privacy capabilities for many energy applications, which has the potential to shape current energy systems \cite{CHENG_bladeicing}. With growing concerns about privacy issues, FL is particularly attractive for constructing a distributed model by pushing machine learning functions to the edge of the energy system when the data reside. Eventually, the data of each energy system is never directly shared with others under the FL framework, while supporting collaborative learning of shared global models, which benefits both grid operators and energy users in terms of network resource savings and enhanced privacy. As such, FL is a powerful alternative to traditional centralized machine learning approaches.

Figure \ref{structure} illustrates the concept of FL, where there are two entities: clients and a central server. The local models can be the end users of an energy system, and the central server is used to aggregate the local models trained by the clients. The general FL process comprises the following three steps:
\begin{figure} [htp]
        \centering
		\includegraphics[width=0.45\textwidth]{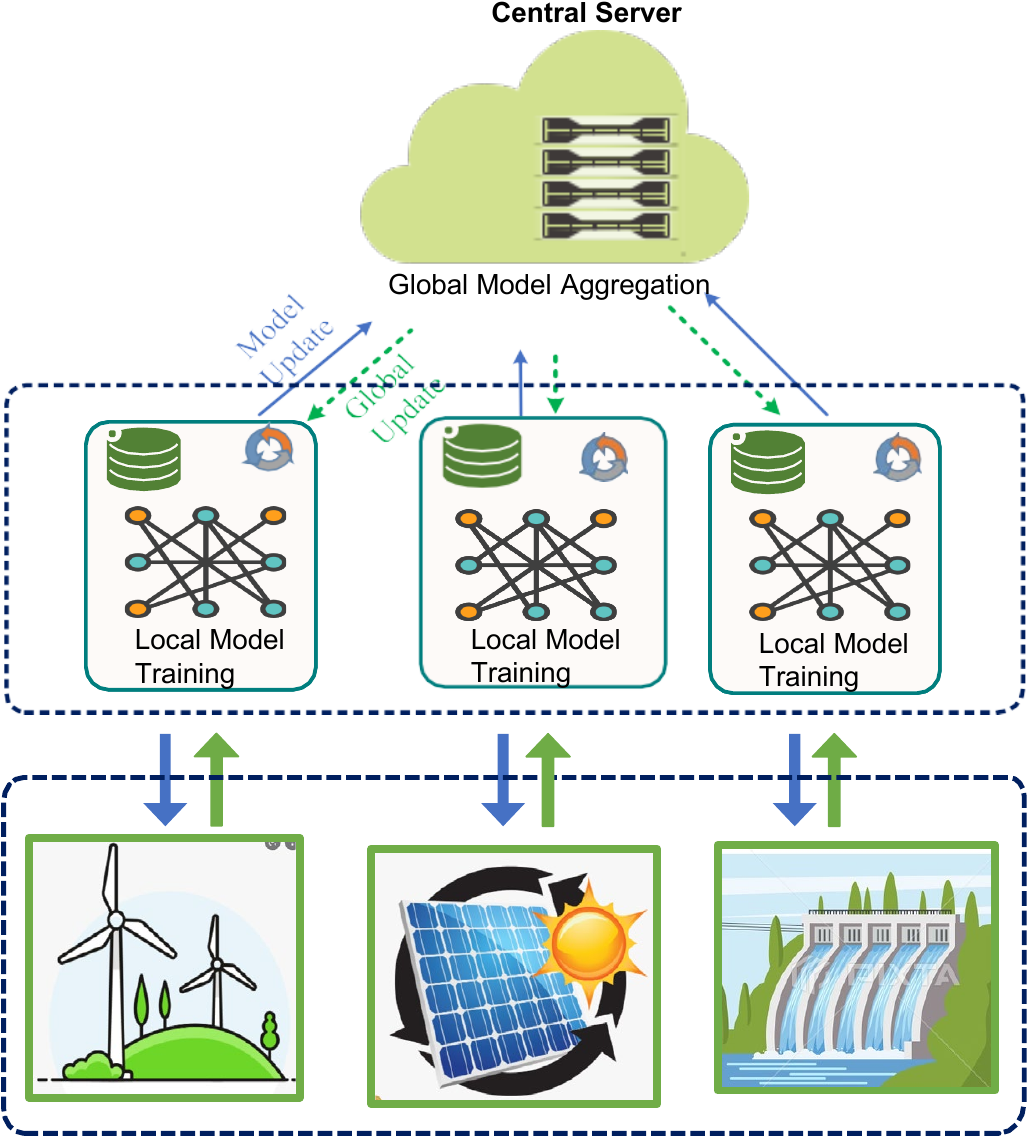}
		\vspace{-5pt}
        \caption{Overview of federated learning applied in an energy system}
        \label{structure}
        \vspace{-15pt}
\end{figure} 

\begin{enumerate}[leftmargin=*]
    \item \textbf{System initialization and device selection}: Most recent work assumes that all clients participate in each training round in the FL process. However, in each training round, only a certain number of clients are selected in practice. 
    
    \item \textbf{Local training}: When training starts, the central server initializes a global model and distributes it to all selected clients. The clients $i$ then start local training using their own local data $d_i$ and calculate the gradient $\omega_i$ by minimizing a loss function $F(\omega_i)$: 
    \begin{equation}
        \omega_i^* = arg min F(\omega_i)
    \end{equation} 
    
    Note that the loss function varies between different FL algorithms depending on the relationship between the input and output pairs. 
    
    \item \textbf{Model aggregation}: After collecting all model updates from the clients, the server will calculate a new global model as :
    \begin{equation}
        \omega_S = \frac{1}{\sum_{i \in N} |d_i|} \sum_{i=1}^N |d_i| w_{i}
    \end{equation} 
    where $\omega_S$ represents the global model. After deriving the global model on the server, the server broadcasts the new $\omega_S$ model to all clients in order to update the old client models. The whole learning process ends when the desired accuracy is reached. 
    
\end{enumerate}

\subsection{Types}
%Reviewing the recent progress of FL algorithm used in the energy domain, we divide FL into two dimensions: data partition and network structure. 
Based on our review of recent advances in FL in the energy domain, we classify its types from the data partitioning and network structure perspectives.

\subsubsection{Data partition}
Depending on how the training data is distributed in the sample space and feature space, FL can be classified into three categories: horizontal FL, vertical FL and federated transfer learning.

\begin{itemize}[leftmargin=*]
    \item \textbf{Horizontal FL}: In horizontal FL, the training data of the clients have the same feature space but different sample spaces. Due to this settings, clients can use the same machine learning model for jointly training the global FL model. To ensure security, the local gradient of the clients usually needs to be protected by encryption techniques \cite{nguyen2021federated}. 
    
    \item \textbf{Vertical FL}: Vertical FL emphasizes shared global model learning in a network of clients with the same sample space and different data feature spaces \cite{feng2020multi}. In vertical FL, a specific alignment method should be designed to collect overlapping data samples from clients. These samples are combined to train a general machine learning model using encryption techniques.
    
    \item \textbf{Federated transfer learning}: Federated transfer learning is designed with the aim of extending the sample space in the vertical FL architecture to more learning clients for data with different sample spaces and different feature spaces \cite{liu2020secure}. In the concept of federated transfer learning, different feature spaces should be transformed into the same representation to aggregate data from multiple clients. To protect data privacy, encryption techniques are also usually widely used. 

\end{itemize}

\subsubsection{Network structure}
Depending on the network topology, FL can be divided into two categories: central server-based FL and distributed FL. 

\begin{itemize}[leftmargin=*]
    \item \textbf{Central server-based FL}: Central server-based FL is one of the most popular FL architectures. Fig. \ref{structure} is a typical structure of the central server-based FL. In this type of FL, there is a central server and a set of clients that execute the FL model. During a training round, all clients participate in the training of the network model in parallel using their own data. All clients then transmit the trained parameters to the central server, which aggregates these parameters using a weighted average algorithm such as FedAvg. Then, the computed global model is sent back to all clients for the next training iteration. At the end of the training process, each client receives an identical global model and its personalized model. This type of FL is highly dependent on the server model to coordinate the aggregation and distribute model updates.  
    
    \item \textbf{Distributed FL}: Distributed FL is a network topology without any central server to coordinate the training process. Instead, all clients are connected in a peer-to-peer (P2P) manner to perform AI training \cite{cheng2022blockchain}. In this way, in each round of communication, clients are also trained locally on their own data. Each client then implements model aggregation using model updates received from neighboring clients via P2P communication to agree on global updates. A distributed FL is designed to completely or partially replace central server-based FL when communication with the server is unavailable or when the network topology is highly scalable. Due to its contemporary features, a distributed FL can be integrated with P2P-based communication technologies such as blockchain to build a decentralized FL system. In this way, distributed FL clients can communicate through a blockchain ledger where model updates can be offloaded to the blockchain for secure model exchange and aggregation. 
\end{itemize}

\subsection{Aggregation algorithms}
Aggregation algorithm plays a key role in FL \cite{mothukuri2021survey}. It incorporates the updates of all local models from all participating clients. The aim of the aggregation algorithm is to: 1) enhance privacy from local model updates; 2) preserve communication bandwidth; 3) facilitate asynchronous updates from clients. As the core of FL, the implementation of the aggregation algorithm varies according to the pre-defined needs. The current most widely used aggregation algorithms can be summarized as follows.

\begin{itemize}[leftmargin=*]
\item  \textbf{FedAvg}: This algorithm was originally proposed by Google. In FedAvg, the central server as the coordinator initiates the FL training process by sharing global parameters or global models with each selected client. Each selected client uses local data to train a global model and shares the weights of the trained model with the central server. The update of the global model is calculated by averaging the local models collected on the central server \cite{mcmahan2017communication}.

\item \textbf{FedProx}: This algorithm is an improved version of FedAvg, which is proposed to address heterogeneity in FL \cite{li2020federated}. FedProx takes into account variations in computing power and different factors in devices participating in FL training rounds. FedProx also introduces a proximal term to handle inconsistencies in local updates. The results of the experiment \cite{li2020federated} indicate that FedProx can achieve positive results in heterogeneous settings.

\item \textbf{FedMA}: This algorithm is used to build a shared model to aggregate CNN and LSTM models in FL \cite{wang2020federated}. FedMA averages the model on the central server through hierarchical matching and averaging hidden elements, such as neurons and channels in the neural network. As shown in the experimental results, FedMA works well on heterogeneous clients, surpassing FedAvg and FedProx in several rounds of training.
\end{itemize}

\section{Use cases}
\label{sec:usecases}
 % Electrification plays a key role in the whole energy landscape as the most efficient and environmental friendly final energy form with higher controllability thanks to the advances on many technologies, such as materials, power electronics, data analysis and more. A revolution of In this survey, we will first focus on the FL application on the smart As a particular data-driven model, we will first introduce the application in a use case perspectives, as this is the natural way to bring data to value.  the use case is of high importance to bring the value to   This section we will focus on summarizing the applications of the FL on the smart grid domain to analysis the state of the .  
 %Electrification plays a key role in the whole energy landscape as the most efficient and environmental friendly final energy form with higher controllability thanks to the advances on many technologies, such as materials, power electronics, information and communication technology (ICT) and more. In this section, we will focus on the FL application for the smart grid mainly. As a particular data-driven model, we will first introduce the FL application in a use case perspectives, and detail analysis of the current application based on the publication collected will be followed based on the both the scenarios and technology differences, to give comprehensive overview of the current research and help identify the problems and potencials which be will covered in Section~\ref{sec:challenges}. 
   %
 Electrification plays a key role in the whole energy landscape as the most efficient and environment-friendly final energy form with higher controllability due to advances in many technologies, such as materials, power electronics, information and communication technology (ICT), etc. In this section, we will focus on FL applications in energy (most of which are in the smart grid). As a particular data-driven model, technology innovation will only bring value when it fits into a particular use case. Therefore, we will first briefly introduce the general conditions of the smart grid domain, which is under dramatic changes. The main players that shape the landscape of the smart grid will be identified.  Next, emerging FL applications based on the current literature review will be summarized from a use case perspective, along with their particular technology contributions. Special attention is paid to the aggregation mechanism for each FL application, as different use cases usually need customized FL frameworks with specific aggregation models. 
\begin{figure} [htp]
        \centering
		\includegraphics[width=0.45\textwidth]{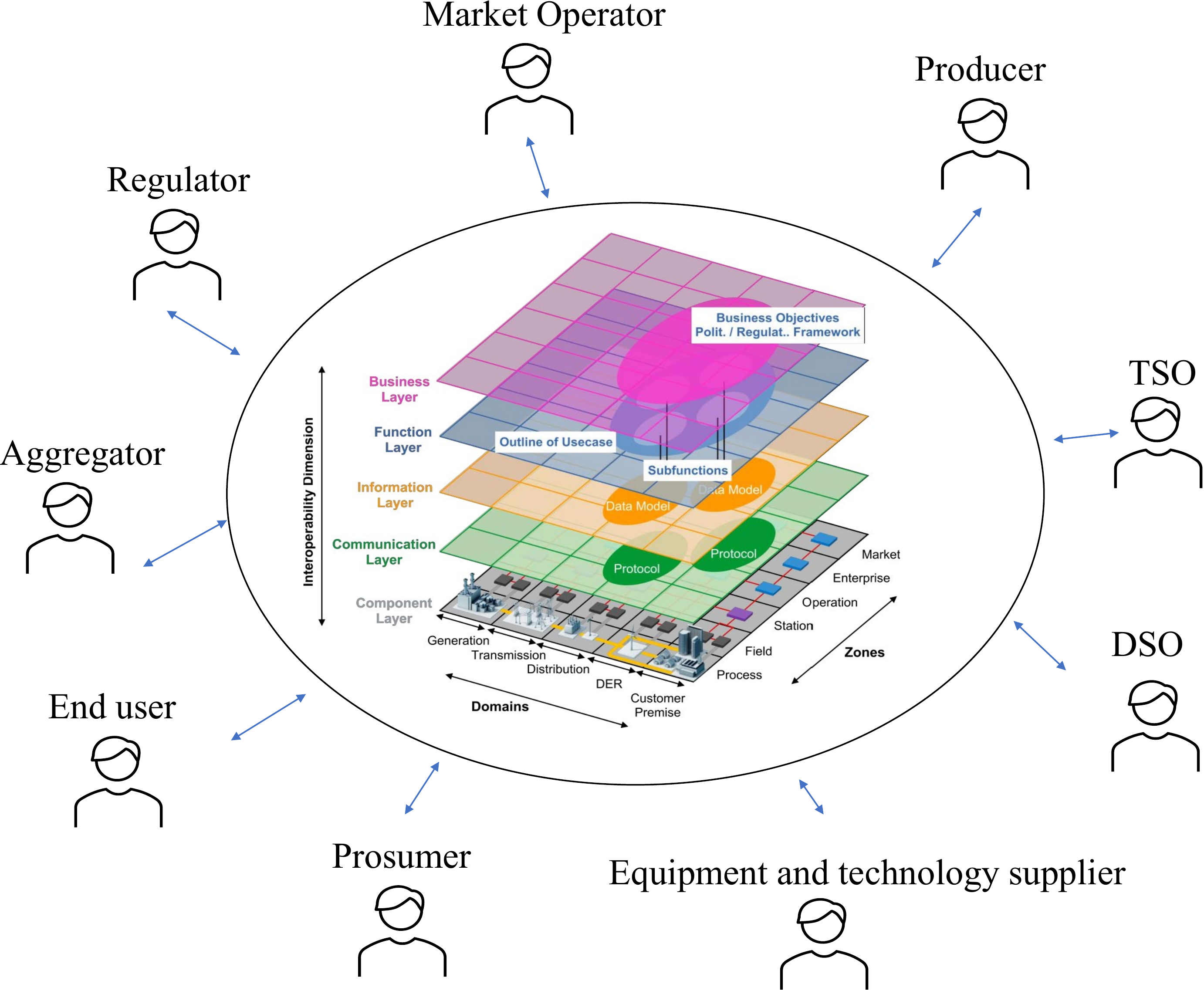}
        \caption{Stakeholders in the Smart Grid}
        \label{stakeholders}
        \vspace{-15pt}
\end{figure} 
      
\subsection{Smart Grid landscape in the IoE and its stakeholders}
The power system witnesses radical revolutions featured by the ``3D"--decolonization, decentralization, digitization, and it renders the system into a smart grid.  Traditional stakeholders in the power sector are also changing due to the structural change from a vertical transmission of the energy starting from large power plants
to end users through transmission lines, to a system which is networked with distributed energy resources. At the same time, more and more new players are entering the energy sector with an increasing deregulated and competitive electricity market. Among these new players, the interests of the prosumers who both consume and generate energy with distributed generations are becoming increasingly non-trivial.
    
As electricity continues to replace other energy forms as the final energy usage type, the interaction between the power sector and other energy sectors becomes more significant. It is no exaggeration to say that the smart grid will be the central enabler in the ecosystem of IoE, with many emerging technologies and applications, such as microgrids as the energy hubs, data analysis, multi-energy flow energy management, etc\cite{li2020defining, li2019data, li2016power}. Fig.~\ref{stakeholders} depicts the current smart grid landscape that engages with different stakeholders in dimensions including interoperability, domains, and zones.  
      
\subsection{Use case summary}
FL is an emerging research field in the energy sector. From a data analysis point of view, applications can be divided into analysis, prediction, and decision making (control and optimization). In this paper, we review the FL use cases for energy systems and summarize in Table~\ref{Tab:summary} according to the use scenarios, the stakeholders benefited, challenges, FL types, and the aggregation feature of FL.

\begin{table*}[t!]
\centering
\arrayrulecolor{black}
\caption{Summary of federated learning for energy systems}
\begin{tabular}{p{0.25cm}p{4.8cm}p{2cm}p{3cm}p{2.6cm}p{3cm}}

\arrayrulecolor{black}\hline
Ref.&Scenarios & Stakeholder & Challenges & FL type    &        Aggregation  feature                                                                                                          \\ 
\arrayrulecolor{black}\hline
\rowcolor[HTML]{ECF4FF}  \hypersetup{citecolor=blue}\cite{linPrivacyPreservingHouseholdCharacteristic2022}&Household characteristic identification, such as income level, flat type, floor area, number and type of applicance and number of residence                        & Retailers  &  Data ownership,
privacy preservation & Horizontal, central server-based FL & Asynchronous stochastic gradient descent with delay compensation (ASGD-DC)\\
\hypersetup{citecolor=blue} \cite{leeFederatedReinforcementLearning2022}& Energy management of multiple smart homes equipped with DER in a system with one home energy management and several local home energy systems.  & Aggregator of LHEMSs & computation burden, timeliness/real time,   privacy preserve & FRL, horizontal & FedSGD algorithm
\\
\rowcolor[HTML]{ECF4FF} \hypersetup{citecolor=blue}\cite{gaoDecentralizedFederatedLearning2021}& Residential   building load forecasting with data from IoT devices                                                                                                     & End user(smart home owner)                                            & data access, privacy preserve,   cloud service cost                            & Horizontal, decentralized FL           & Decentralized, global server free                                                                          \\
\hypersetup{citecolor=blue}\cite{fekriDistributedLoadForecasting2021 }&Load forecasting using smart meter   data                                                                                                                                 & utility                                                               & computation burden, data   ownership, privacy preserve                         & horizontal,   Central server based FL  & FedAVG                                                                                                     \\
\rowcolor[HTML]{ECF4FF}\hypersetup{citecolor=blue}\cite{taikEMPOWERINGPROSUMERCOMMUNITIES2021} & Market   participate and energy exchange with prosumer community   group                                                                                                 & Aggregator-prosumer community   groups, end user                      & privacy preserve, data access                                                  & Horizontal,   central server based FL  & Averaging                                                                                                  \\
\hypersetup{citecolor=blue}\cite{wangElectricityConsumerCharacteristics2021}& Identificaiton   of socio-demographic characteristics of electricity consumers from smart   meter   data                                                            & load forecasting using smart meter data                               & privacy preserve                                                               & Horizontal,   central server based FL  & Model averaging                                                                                            \\
%Household   Characteristic Identification, such as income level, flat type, floor area,   number and type of applicance and number of   residence\hypersetup{citecolor=blue}\cite{linPrivacyPreservingHouseholdCharacteristic2022}              & retail company                                                        & data ownership， privacy preserve & horizontal, Central server based   FL  & asynchronous stochastic gradient   descent with delay compensation (ASGD-DC)                               \\
\rowcolor[HTML]{ECF4FF} \hypersetup{citecolor=blue}\cite{bahramiDeepReinforcementLearning2021}& Demand   response through load control of multiple households to reduce peak load and user  cost & Aggregator, end user(residential) & privacy preserve &  Horizontal, central server based FL,   RL & decentralized algorithm based on   proximal Jacobian alternating direction method of multipliers (PJ-ADMM) \\
\hypersetup{citecolor=blue}\cite{linPrivacyPreservingFederatedLearning2022}&Behind-the-meter   solar disaggregation from the net load to  the solar generation profile at   the community   level                                             & Utility                                                               & Privacy preserve                                                               & Horizontal, central server based   FL  & A layerwise parameter aggregation   strategy                                                               \\
\rowcolor[HTML]{ECF4FF}\hypersetup{citecolor=blue}\cite{leeFederatedReinforcementLearning2022}& Home   Energy management System(HEMS) of multiple smart homes equipped with DER.                                                                                          & Aggregato (HEMS that consists of a   single GS and N LHEMSs)           & computation burden, real time,   privacy preserve                              & FRL, horizontal                        & FedSGD algorithm                                                                                           \\
\hypersetup{citecolor=blue}\cite{cheng2022class}& Blade icing detection for wind turbines in the wind   farms                                                                                                                                      & wind farm owners                                                      & data ownership, computation   burden                                           & Horizontal, central server based   FL  & Differentiated computational capabilities in   server and client models                                       \\
\rowcolor[HTML]{ECF4FF}\hypersetup{citecolor=blue}\cite{liPrivacypreservingSpatiotemporalScenario2021} &Spatiotemporal   scenario generation of renewable energies                                                                                                       & Utility                                                               & privacy preserve                                                               & Horizontal, central server based   FL  & Integrating federated learning and least   square generative adversarial networks                          \\
 \hypersetup{citecolor=blue}\cite{thorgeirssonProbabilisticPredictionEnergy2021}&Probabilistic   prediction of energy demand and driving range for electric vehicles to deal   with limited driving range in a sparse charging infrastructure    & EV owner                                                              & privacy preserve, computation and communication overhead                       & Horizontal, central server based   FL  & Federated Learning With   Clustering, FedAvg and FedAG                                                     \\
\rowcolor[HTML]{ECF4FF}\hypersetup{citecolor=blue}\cite{wenFedDetectNovelPrivacyPreserving2021}& Energy   theft detection in smart grid                                                                                                                                  & Utility                                                               & privacy preserve                                                               & Horizontal, central server based   FL  & FL with homomorphic encryption                                                                             \\
\hypersetup{citecolor=blue}\cite{saviShortTermEnergyConsumption2021}& Residential short-term energy consumption forecasting                                                                                                                     & Utility                                                               & privacy preserve                                                               & Horizontal, central server based   FL  & FedAVG                                                                                                     \\
\hypersetup{citecolor=blue}\cite{wangFedNILMFederatedLearningbased2021}& Non-intrusive   load monitoring                                                                                                                                          & Utility                                                               & data access, privacy preserve                                                  & Horizontal, central server based   FL  & FedAVG                                                                                                     \\
\rowcolor[HTML]{ECF4FF} \hypersetup{citecolor=blue}\cite{saputraEnergyDemandPrediction2019}& Predicting energy demand for electric vehicle (EV) networks                                                                                                                & Grid operator                                                         & privacy preserve                                                               & Horizontal, central server based   FL  & Clustering-based FL                                                                                        \\
\hypersetup{citecolor=blue}\cite{saterFederatedLearningApproach2020}& Anomaly Detection in Smart Buildings                                                                                                                                      & End user-the owner of the smart   buildings                           & computation   burden,timeliness,                                               & Horizontal, central server based   FL  & FL with  encryption(PySyft)                                                                                \\
\rowcolor[HTML]{ECF4FF} \hypersetup{citecolor=blue}\cite{linIncentiveEdgebasedFederated2022}& False data injection attack detection on power grid state estimation                                                                                                      & Utility-power system state owner,   detection service provider (DSP) & data access, privacy preserve,   data ownership                                & Horizontal, central server based   FL  & Federated learning with incentive                                        
\\
\hline
\end{tabular}
\label{Tab:summary}
\vspace{-10pt}
\end{table*}

% as this is the natural way to bring data to value.  the use case is of high importance to bring the value to   This section we will focus on summarizing the applications of the FL on the smart grid domain to analysis the state of the .   

\section{Challenges and opportunities}
\label{sec:challenges}
FL is a new paradigm of distributed learning across different devices, data centers, and geographic locations. It is also a multidisciplinary learning framework that involves several disciplines, including cryptography, databases, and machine learning. Among many others, the challenges and opportunities of FL are summarized as follows: 1) How to improve the efficiency of the communication between the central server and the clients, which mainly involves how to reduce the number of communications and compress the data. 2) How to protect the security of the data on the communication during the model training process. In this regard, data encryption techniques can be applied, including homomorphic encryption and differential privacy. 3) How to partition the data on the client side to achieve optimal parallel learning efficiency. Existing studies focus on the application of horizontal and vertical data partitioning strategies. 4) How to deal with non-identical distribution problems of the training data. These can be problems of skewed feature distribution and/or skewed label distribution. 5) How to improve learning efficiency, which mainly consists of studying how to optimize the machine learning algorithm and accelerate model convergence. 6) How to achieve multitask learning and personalized learning in a federated setting. Since clients may have different models, settings, and devices, this heterogeneity determines the challenges of balancing parallel tasks, thus providing personalized learning opportunities. 7) How to design machine learning algorithms and tune hyperparameters in an FL framework.

The above are the main challenges and opportunities for FL technology itself, which has been the subject of many research efforts. As a result, this has led to a large number of review articles; in this paper, we instead have focused on the other aspect that we think is most interesting, but has not yet been examined, namely, the global model generation algorithms. Aggregation algorithms are directly related to the performance of the final model, such as generality. In addition, we believe that FL should be fully decentralized or support peer-to-peer distributed learning, similar to blockchain technology where learning is consensus-based. Finally, due to the complexity of implementing FL technology and the applications it supports, we believe it would be beneficial to integrate visualization into an FL framework, which can greatly lower the learning curve and improve the interpretability of machine learning models implemented in an FL framework.

FL has a wide range of potential in the field of energy, which can be summarized as follows. First, it is often difficult to create powerful data-driven models for renewable energy prediction, mainly because of inherent stochasticity. For example, wind power generation is affected by wind speed, and solar photovoltaic energy is affected by the intensity and timing of solar exposure. Training a forecasting model with high accuracy and generality requires different data sources, which may be from power plants at different geographical locations. However, these data may not be available due to data ownership or commercial competition, in which case FL becomes essential to create a collaborative prediction model. Second, in the residential building sector, FL can help protect the privacy of customer information. For example, in Europe, the residential electricity supply is often operated by different distributors, each of which has its own customers. In order to create data-driven models from energy consumption and customer information, FL can address the model in a non-sharing environment with data privacy guarantee. This makes it possible to provide personalized energy services, such as energy-efficiency product recommendations, demand-response programs, and customized energy management solutions. Although research has been conducted on personalized recommendations based on FL in other fields, its applications in the energy sector remain an open problem that is interesting to investigate.  Finally, energy involves many distributed optimization problems, including production, transportation, and distribution. The advantage of applying federated-based optimization is that it supports heterogeneous models, i.e., the model structure does not need to be consistent across different nodes. Therefore, an ensemble-like heterogeneous distributed optimization model can be established to help with decision making regarding energy planning or supply.

We are aware that energy is a very broad concept that has many subdomains. For example, depending on the types of energy, it can be divided into water, electricity, gas, heat, etc., each with its own life cycle for an energy system. In addition, there are many other energy-related research areas, including market, price, regulation, and policies. Therefore, FL has a wide potential in energy.

\section{Conclusion and future work}
\label{sec:con}
FL is an emerging computing paradigm. In this paper, we presented a systematic review of FL and its applications in the energy sector. We first introduced the concept of FL, the technical advances especially about the research of solutions, network structures, and global model generation algorithms. Then, we investigated various solutions and applications of FL in energy, mainly involving demand-response programs, user characteristics identification, customer segmentation, energy prediction, and optimization. We also discussed the technical challenges and opportunities of the FL technology itself and enumerated its opportunities in the energy sector. We hope that this survey will help researchers refine FL techniques and provide useful reference material for researchers in the energy field.

In the future work, we will improve our review in a more systematic way and update the latest research advances on FL and its applications in energy. We also intend to develop an FL benchmarking and evaluation tool, as no such tool is available to date.

\bibliographystyle{IEEEtran}
\bibliography{IEEEabrv, Reference}

\end{document}